\def\broj{23}
\def\vol{1}
\def\str{1--9}
\def\udk{523.985.3-355}
\def\vrsta{Original scientific paper}

\def\gore{j. murak\"{o}zy et al.: sunspots and torsional waves}

\documentclass{hob}  
\usepackage{epsfig}

\begin{document}

\title{Search for possible connections of sunspot features and torsional waves}

\author{\normalsize J. MURAK\"{O}ZY, G. MEZ\H{O}, A. LUDM\'{A}NY, \\
         \normalsize  L. GY\H{O}RI \vspace{2mm} \\
        \it Heliophysical Observatory H--4010 Debrecen, P.O.B.30. Hungary,\\}

\maketitle

\begin{abstract}
The torsional oscillation is a well established observational fact
and there are theoretical attempts for its description but no final solution 
has yet been accepted. One of the possible candidates for its cause is
the presence of sunspots modifying the streaming conditions. The present work 
focuses on the temporally varying latitudinal distribution of several sunspot
features, such as the spot sizes and spot numbers. These features are different
faces of the butterfly diagram. In fact some weak spatial correlations can 
be recognized.

\end{abstract}

\keywords{Sun - torsional waves - sunspot features} 

\section{Introduction}
\large
The torsional oscillation was discovered by Howard and LaBonte {1980} as a travelling wave
pattern superposed on the differential rotation. The waves propagate from the poles toward
the equator and they consist of prograde and retrograde zones with respect to the mean 
differential rotation profile, the amplitude of the deviations is about 7 m/sec. Two waves 
(two prograde and retrograde belts) coexist in both hemispheres. The feature was absolutely
unexpected at the time of discovery and several attempts have been made to check and interpret it. 
The empirical studies basically confirmed the existence of the phenomenon by using either 
surface measurements (Ulrich, 2001) or subsurface detection by GONG and MDI data 
(Howe et al., 2000; Komm et al 2001; Zhao and Kosovichev, 2004). As a result of these 
investigations the torsional oscillation can be regarded to be a persistent feature which 
extends down to about 0.92 $R_{\odot}$.

The theoretical approaches were motivated by the similarity of the equatorward migrations of 
the shearing belts and the activity belts. This suggested that the activity cycle 
(in particular the Sp\"orer's law) must have something to do with the torsional oscillation.
The first description was published by Yoshimura (1981) who suggested a mechanism
driven by a Lorentz force wave as a by-product of the dynamo wave. Recent models take into 
account the presence of the sunspots. In the model of Petrovay and Forgcs-Dajka (2002)
the sunspots modify the turbulent viscosity in the convective zone which leads to the modulation
of the differential rotation. In the model proposed by Spruit (2003) the sunspots exert
a cooling effect on the surface and this temperature variation results in geostrophic flows 
which would drive the torsional oscillation. 

The present work was motivated by these recent works which suggested a possible connection between
the sunspots and the torsional oscillation phenomenon. We would like to find any spatial
correlation between the torsional wave and any sunspot feature. Earlier works also indicated
some spatial connections but e.g. LaBonte and Howard (1982) averaged the latitudinal magnetic 
activity distribution for a longer time whereas Zhao and Kosovichev (2004) only indicated 
the location of the activity belt with no distribution information. 

The aim of the present work is to follow the temporal variation of the latitudinal distribution 
of sunspot parameters in comparison with the migration of the torsional waves. As a first 
attempt three parameters were chosen to scrutiny. Two of them may be trivial: the number and the 
total area of sunspots. The third parameter, the mean number of sunspots within the groups 
characterizes the complexity of the sunspot groups, because it can also be expected that the more 
complex is a sunspot group the more efficient is its interconnection with the local velocity fields.

\section{Observational Data}

The sunspot data were taken from the most detailed sunspot database, the Debrecen Photoheliographic
Data (DPD). This is the only material which contains the position and area data for each observable
spot, even for the smallest ones, for each day. The temporal coverage of the DPD was partial at the
time of this work so, as a first attempt, we restricted the study to the years 1986-1989 (Gy\H ori et al, 
1996). The latitudinal distributions were determined in such a way that 1 degree latitudinal stripes 
were considered, and all mentioned parameters (total number and total area of spots, as well as the 
mean number of spots per goups) were added up for all stripes and three months. A total amount 
XXX spots were taken into account in the given period.

To compare the obtained distributions with the torsional pattern one has to determine the latitudinal
location of the torsional wave i.e. the latitudes of prograde/retrograde belts and the shearing zones.
For this period the most suitable torsional data were provided by Ulrich (2001). In his figure 1. 
the shearing zones were reasonably well recognizable in this period. Figure 1. shows the period 1986-89
taken from Ulrich (2001) and the most probable lines of the shear zones (dark regions indicate the 
prograde belts). In our further figures these lines were adopted to mark the migrating shear zones.

\begin{figure}[ht]
 \begin{center}
\epsfig{file=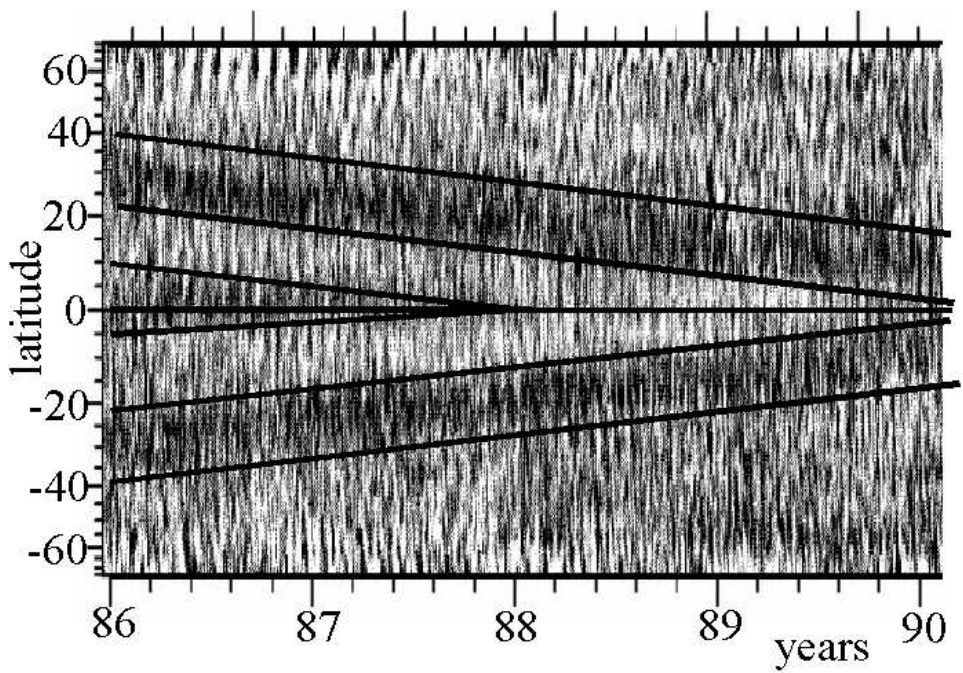,width=12.5cm}
 \caption{Locations of the shear zones superposed on the velocity map of Ulrich (2001).}
 \end{center}
\end{figure}

\section{Sunspot features}

The most obvious candidate may be the number of sunspots at a certain latitude. The numbers of all 
spots have been added up in 1 degree wide stripes and 3-month periods in such a way that each sunspot
group was taken into account at that time when it contained the largest number of spots. The resulting
distributions were plotted onto the plot of the migrating zones, see Figure 2, where the temporal 
dimension has been streched in order to minimize the overlap of the distribution curves. In this 
approach the sizes of the spots were omitted, only the amount of the magnetix flux tubes was 
considered regardless of their strengths.

\begin{figure}[ht]
 \begin{center}
\epsfig{file=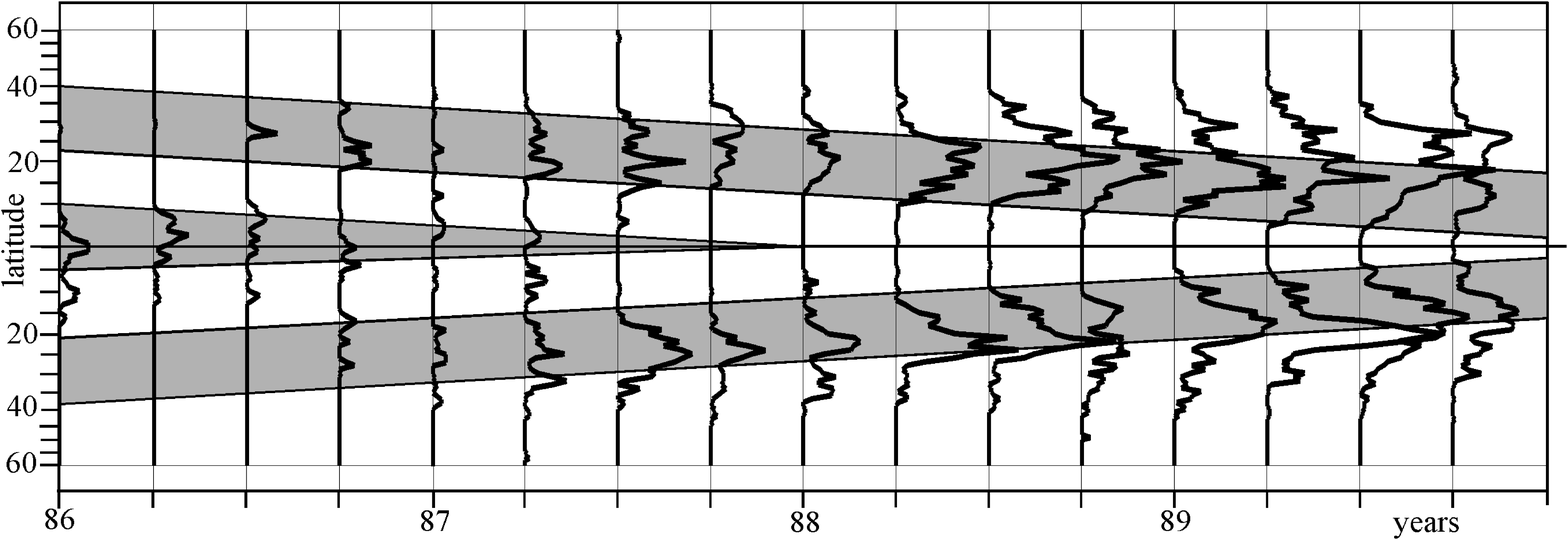,width=12.5cm}
 \caption{Latitudinal distributions of the number of sunspots in comparison with the torsional belts.}
 \end{center}
\end{figure}

The next possible candidate is the total area of the spots (Figure 3). The procedure was the same as
in the case of the sunspot numbers, all area data were added up by $1^{\circ}$ latitude stripes. Sunspots
were taken into account at that time when the total areas of their groups were the largest during their
passage through the solar disc. In this approach the total strength was considered regardless of the size 
distribution.

\begin{figure}[ht]
 \begin{center}
\epsfig{file=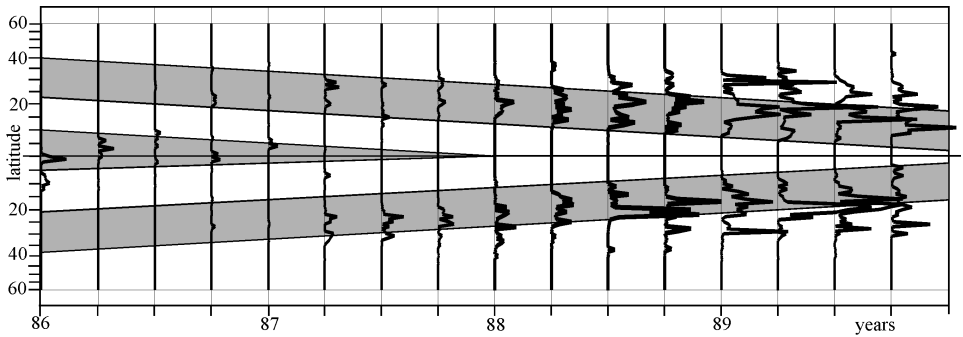,width=12.5cm}
 \caption{Latitudinal distributions of the total area of sunspots in comparison with the torsional belts.}
 \end{center}
\end{figure}

The third candidate was chosen to check the assumption that in case of a certain sunspot-torsion
interconnection the more complex sunspot groups could exert more efficient impacts on the ambient
flows. The number of spots within the groups has been averaged for the $1^{\circ}$ latitude stripes
and 3-month periods, the groups were considered at the time of their largest extensions. The resulting
distributions are displayed in Figure 4.

\begin{figure}[ht]
 \begin{center}
\epsfig{file=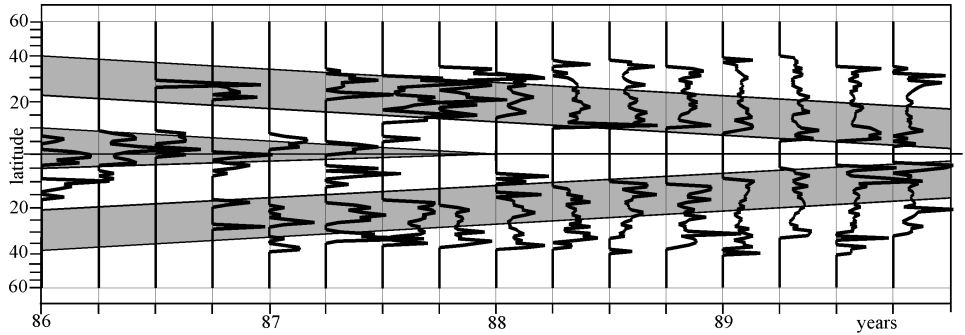,width=12.5cm}
 \caption{Latitudinal distributions of the mean number of sunspots per groups in comparison with the torsional belts.}
 \end{center}
\end{figure}

\section{Discussion}

It is remarkable in the Figures 2. and 3. that the distributions of both the spot number and area are so positioned
that the peaks of the curves (if they have unambiguous peaks at all) are close to the shear zones but the bulges of the
distributions are mostly situated in the faster belts. The curves of the number and area are mostly different,
as was expected, but the mentioned character seems to be overwhelming on both cases. This behaviour may be the
signature of a really functioning interconnection between the torsional pattern and the magnetic flux ropes.

As for the complexity of the groups, no such trend can be recognized so this feature (the number of spots per groups)
is apparently unimportant from this point of view.

The above features are the first (preliminary) results of our project and they seem to be encouraging. In the 
following works we are intended to extend the temporal domain of the study and to include also some 
further possible sunspot parameters to reveal the nature of this interaction.

\section{Acknowledgements}

The present work was supported by the grants OTKA T 37725 and ESA PECS No.98017. One of the authors 
(A.L.) expresses his gratitude for the kind invitation and hospitality of the Hvar Observatory.

\section*{References}
\begin{itemize}
\small
\itemsep -3pt
\itemindent -20pt

\item[] Gy\H ori, Baranyi, T., Csepura, G., Gerlei, O., Ludm\'any A. 1996, 
  Debrecen Photoheliographic Data for the year 1986 Publ. Debrecen Obs. Heliogr. Ser. 10, 1-61.
\item[] Howard, R., LaBonte, B. J.: 1980, {\it Astrophys. J.} {\bf 239}, L33-L36.
\item[] Howe, R. {\it et al.}: 2000,{\it Astrophys. J.} {\bf 533}, L163-L166.
\item[] Komm R.W., Hill F., Howe R., 2001, {\it Astrophys. J.} {\bf 558} 428-441.
\item[] LaBonte, B. J.,Howard R.: 1982,{\it Solar Physics} {\bf 75}, 161-178.
\item[] Petrovay, K., Forg\'acs - Dajka E.,: 2002, {\it Solar Physics} {\bf 205}, 39-52.
\item[] Spruit, H. C.: 2003,{\it Solar Physics} {\bf 213}, 1-21.
\item[] Ulrich, R. K.: 2001,{\it Astrophys. J.} {\bf 560}, 466-475.
\item[] Yoshimura, H. {\it Astrophys. J.} {\bf 247} 1102-1112.
\item[] Zhao, J., Kosovichev, A. G.: 2004,{\it Astrophys. J.} {\bf 603}, 776-784.
\end{itemize}


\end{document}